\documentclass[epj]{webofc}
\usepackage[varg]{txfonts}   
\usepackage[version=3]{mhchem}
\woctitle{EPOV 2012}
\begin{document}
\title{VUV-absorption cross section of \ce{CO2} at high temperatures and impact on exoplanet atmospheres}
%
%

\author{Olivia Venot\inst{1,2,3}\fnsep\thanks{\email{olivia.venot@ster.kuleuven.be}} \and
        Nicolas Fray\inst{4} \and
        Yves B\'{e}nilan\inst{4} \and
        Marie-Claire Gazeau\inst{4} \and  
        Eric H\'{e}brard\inst{1,2} \and
        Gwenaelle Larcher \inst{4} \and
        Martin Schwell  \inst{4} \and
        Michel Dobrijevic\inst{1,2} \and
        Franck Selsis\inst{1,2}
}

\institute{Univ. Bordeaux, LAB, UMR 5804, F-33270, Floirac, France
\and
           CNRS, LAB, UMR 5804, F-33270, Floirac, France 
\and
           Instituut voor Sterrenkunde, Katholieke Universiteit Leuven, Celestijnenlaan 200D, 3001 Leuven, Belgium
\and
Laboratoire Interuniversitaire des Syst\`{e}mes Atmosph\'{e}riques, UMR CNRS 7583, Universit\'{e}s Paris Est Cr\'{e}teil (UPEC) et Paris Diderot (UPD), Cr\'{e}teil, France          
          }

\abstract{
Ultraviolet (UV) absorption cross sections are an essential ingredient of photochemical atmosphere models. Exoplanet searches have unveiled a large population of short-period objects with hot atmospheres, very different from what we find in our solar system. Transiting exoplanets whose atmospheres can now be studied by transit spectroscopy receive extremely strong UV fluxes and have typical temperatures ranging from 400 to 2500~K. At these temperatures, UV photolysis cross section data are severely lacking. Our goal is to provide high-temperature absorption cross sections and their temperature dependency for important atmospheric compounds. This study is dedicated to \ce{CO2}, which is observed and photodissociated in exoplanet atmospheres. We performed these measurements for the 115 - 200~nm range at 300, 410, 480, and 550~K. In the 195 - 230~nm range, we worked at seven temperatures between 465 and 800~K. We found that the absorption cross section of \ce{CO2} is very sensitive to temperature, especially above 160~nm. Within the studied range of temperature, the \ce{CO2} cross section can vary by more than two orders of magnitude. This, in particular, makes the absorption of \ce{CO2} significant up to wavelengths as high as 230~nm, while it is negligible above 200~nm at 300~K. 
To investigate the influence of these new data on the photochemistry of exoplanets, we implemented the measured cross section into a 1D photochemical model. The model predicts that accounting for this temperature dependency of \ce{CO2} cross section can affect the computed abundances of \ce{NH3}, \ce{CO2}, and CO by one order of magnitude in the atmospheres of hot Jupiter and hot Neptune.
}
\maketitle
\section{Introduction}
\label{intro}

Exoplanets exhibit a wide variety of mass, radius, orbits, and host stars. Because of observational biases, most known transiting exoplanets are very close to their parent stars and are highly irradiated, implying large UV fluxes and high atmospheric temperatures. The atmosphere of transiting hot Jupiters and hot Neptunes can be studied by spectroscopy at the primary transit \citep{tinetti2007water, tinetti2007infrared, swain2008presence, beaulieu2010water, Tinetti2010, 2011beaulieu} and at the secondary eclipse \citep{swain2009water, swain2009molecular, stevenson2010possible, stevenson2012two}. Photochemistry has an important influence on the atmospheric composition of these exoplanets, from the top of the atmosphere down to 100~mbar \citep{moses2011disequilibrium, line2011thermochemical, venot2012}. For these exoplanets and within this large pressure range, the temperature can vary roughly from 400 to 2500~K. To model correctly the photochemistry of these planets, we need to use absorption cross sections consistent with these temperatures for all the species whose photolysis plays an important role in either the formation/destruction of molecules or in the penetration of the UV flux into the atmosphere. Carbon dioxide (\ce{CO2}) is one of these species. It has been observed in extrasolar giant planet atmospheres \citep{swain2009water, swain2009molecular}, but cross section measurements ($\sigma_{\ce{CO2}}(\lambda,T)$) are extremely sparse above room temperature.

The first experiments dedicated to the determination of absorption cross sections of \ce{CO2} at temperatures different from 298~K were motivated by solar system planetary studies (Mars, Titan, Venus, primitive Earth) so were performed at lower temperatures \cite{Lewis1983297, yoshino96b, parkinson2003, stark2007}.
Some high-temperature measurements have been performed in the past but only at a few wavelengths \cite{Koshi1991519, generalov1963absorption}. These measurements were limited to a narrow range of wavelengths and do not provide complete spectra. Nevertheless, they showed that the absorption of \ce{CO2} increases at high temperature. Spectra between 190 and 355 nm were obtained at very high temperatures (900-4500~K) by \cite{jensen1997ultraviolet, Schulz200282, oehlschlaeger2004ultraviolet} and some of them fitted the strong temperature dependence of $\sigma_{\ce{CO2}}(\lambda,T)$ with an empirical function. 

To our best knowledge, no measurements exist of the absorption cross section of \ce{CO2} between 300 and 900~K in the wavelength range useful for exoplanetary studies ($<$190~nm). We resume here the results that have been published in \cite{venot2013}, that is to say the measurements of the absorption cross section of \ce{CO2} at 300, 410, 480, 550~K between 115 and 200~nm, as well as between 195 and 230~nm at seven temperature values between 465 and 800~K. We also determine a semi-empirical formula to fit the temperature dependence for wavelengths longer than 170~nm. Finally, we study the effect of these new data on the atmospheric composition predicted by a 1D photochemical model of hot exoplanet atmospheres.

\section{Experimental methods}\label{experimental}
We used gaseous \ce{CO2} of 99.995\% purity. Tunable VUV light between 115 and 200~nm was obtained from the synchrotron radiation facility BESSY in Berlin and measurements in the 195 - 230~nm range were performed at the Laboratoire Interuniversitaire des Syst\`{e}mes Atmosph\'{e}riques (LISA) in Cr\'{e}teil, France. In both cases, an oven (Nabertherm) was used to heat the cell to a temperature of 1400~K. A figure of our experimental setup and more details about the experimental methods and the calculation of the absorption cross section are detailed in \cite{venot2013}.

\section{Results and discussion}\label{results}
\subsection{Photoabsorption cross section from 115 nm to 200 nm}

Before heating the gas, we measured ambient temperature (300~K) spectra of \ce{CO2} in order to calibrate and compare it with the previously published data \citep{yoshino96b, parkinson2003, stark2007}; our measurements agree very closely with these measurements with a difference of less than a few percentage points for all wavelengths. Our measurements at room temperature did not go up to 200 nm, so between 170 and 200 nm we use the data of \cite{parkinson2003}.\\

\begin{figure}
\centering
\includegraphics[width=0.8\columnwidth]{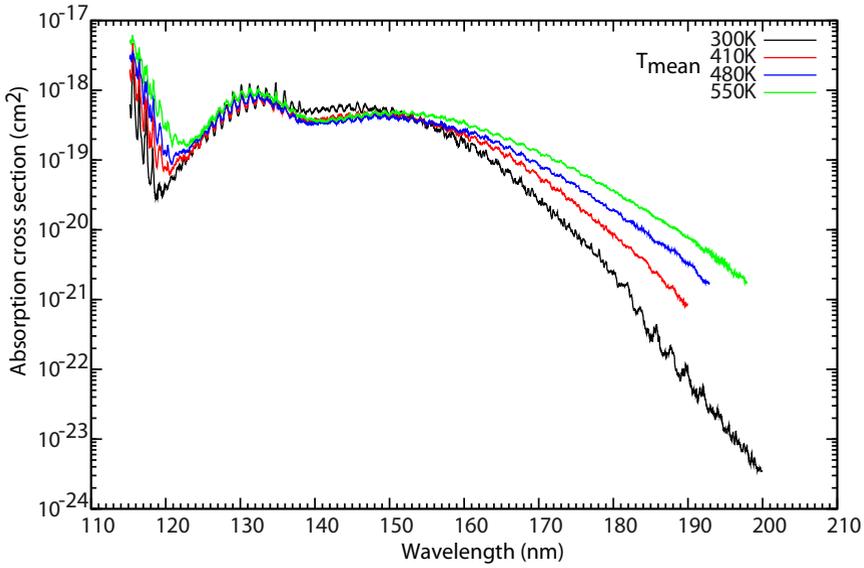}
\caption{Absorption cross section of \ce{CO2} at $T_{mean}$ = 300~K (black), 410~K (red), 480~K (green) and 550~K (blue) for wavelengths between 115 and 200~nm.}
\label{fig:BESSY}
\end{figure}

Then, we measured $\sigma_{\ce{CO2}}(\lambda, T)$ at three different temperatures: 410 ($\pm$~15) K, 480 ($\pm$~25) K, and 550 ($\pm$~30) K. We show these data in Fig.~\ref{fig:BESSY}.
Between 115 and 120~nm we see a change of the cross section which depends on the temperature. At 120~nm, the absorption cross section is ten times higher at 550~K than at 300~K. Slight differences of up to 50\% can be observed between 125 and 140~nm while between 140 and 150~nm differences are minor.
After 160~nm, we clearly observe large differences between the different temperatures. The slope of the cross section varies with the temperature. The higher the temperature is, the less steep is the slope. At 195~nm, there is a factor $\sim$200 between $\sigma_{\ce{CO2}}$($\lambda$,~300~K) and $\sigma_{\ce{CO2}}$($\lambda$,~550~K).\\

\subsection{Photoabsorption cross section from 195 nm to 230 nm}

We measured $\sigma_{\ce{CO2}}(\lambda, T)$ at seven different temperatures: 465~($\pm$~20)~K, 510~($\pm$~25)~K, 560~($\pm$~30)~K, 610~($\pm$~35)~K, 655~($\pm$~45)~K, 750~($\pm$~55)~K and 800~($\pm$~60)~K. As for the cross section at shorter wavelengths, we clearly see the dependence on the temperature in this wavelength range and the increase of the cross section for high temperatures (Fig.~\ref{fig:PARAM}). As we plotted the data obtained previously at shorter wavelengths in this figure, we see good agreement between the two ranges. Especially, we see that $\sigma_{\ce{CO2}}$($\lambda <$~200~nm, 550~K) matches almost perfectly with $\sigma_{\ce{CO2}}$($\lambda >$~195~nm, 560~K).

\subsection{Determination of an empirical law}

\begin{figure}[!ht]
\centering
\includegraphics[width=0.8\columnwidth]{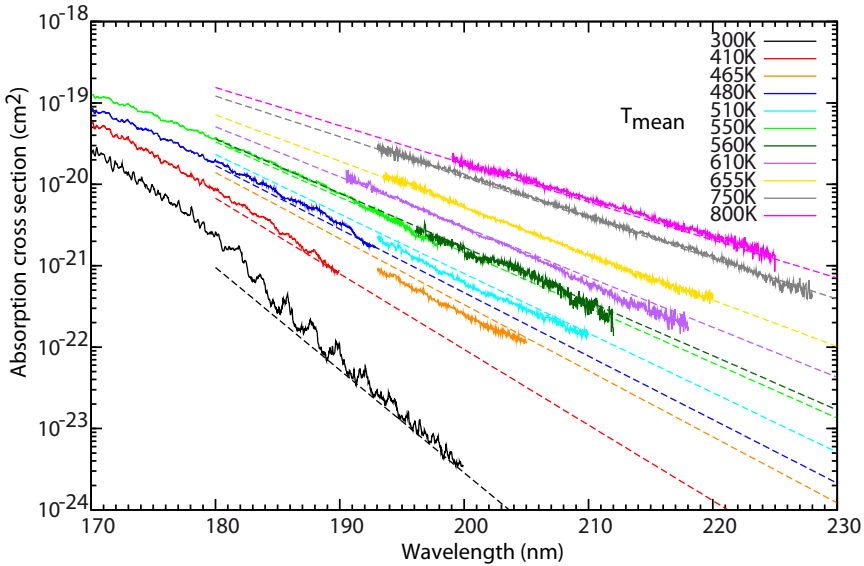}
\caption{Absorption cross section of \ce{CO2} for wavelengths longer than 195~nm at 465~K, 510~K, 560~K, 610~K, 655 K, 750~K, and 800~K, plotted with the cross section at ambient temperature (black) and the absorption cross sections measured at shorter wavelengths and presented in Fig.~\ref{fig:BESSY} (300~K, 410~K, 480~K, and 550~K). The absorption cross sections calculated with Eq.~\ref{eq:formule} are plotted with the same color coding.}
\label{fig:PARAM}
\end{figure}

For wavelengths longer than 170 nm, we parametrize the variation of $\ln(\sigma_{\ce{CO2}}(\lambda,T) \times \frac{1}{Q_v(T)})$ with a linear regression

\begin{equation}\label{eq:formule}
\ln \left(\sigma_{\ce{CO2}}(\lambda,T) \times \frac{1}{Q_v(T)}\right) = a(T) + b(T) \times \lambda
\end{equation}
with $T$ in K and $\lambda$ in nm and where \\
$a(T)= -42.26 + (9593\times 1.44/T)$, \\
$b(T) =  4.82\times10^{-3} - 61.5\times 1.44/T$,\\
and \\
$Q_v (T) = (1- \exp(-667.4\times 1.44/T))^{-2} \times (1-\exp(-1388.2 \times 1.44/T))^{-1} \times (1-\exp(-2449.1\times 1.44/T))^{-1}$\\
is the partition function. Figure \ref{fig:PARAM} compares the absorption cross sections obtained with this calculation to the measurements. We can see that the parametrization is very good.

A discussion about transitions between electronic states and our parametrization can be found in \cite{venot2013}, as well as a discussion about the disagreement between our data and the data of \cite{Schulz200282}.

\section{Application to exoplanets}\label{sec:application}

We investigated the impact of the temperature dependency of the \ce{CO2} cross section on a prototype planet whose characteristics are similar to those of the hot Neptune GJ 436b \cite{butler2004neptune, southworth2010homogeneous}. We chose this planet because the temperature of its upper atmosphere is around $\sim$500~K which corresponds to the highest temperature for which we measured the cross section between 115 and 200 nm. We considered three different types of host stars: an M, a G, and an F star. We used the spectra of GJ 644 (M3V, \cite{segura2005biosignatures}), the Sun (G2V, \cite{gueymard2004}), and HD 128167 (F2V, \cite{segura2003ozone}), scaled to get the bolometric flux received by GJ436b (Figure 9 in \cite{venot2013}).

We used the model described in \cite{venot2012} and the same temperature profile as \cite{line2011thermochemical} calculated by \cite{lewis2010atmospheric}. Although this is not a realistic assumption, we used the same temperature profile for all three host stars (Figure 10 in \cite{venot2013}). To model the vertical mixing, we considered an eddy diffusion coefficient constant $K_{zz}=10^{8}$cm$^2$.s$^{-1}$. Elemental abundances of the atmosphere of this planet are highly uncertain \cite{stevenson2010possible, madhusudhan2011highGJ436b}. So we assumed a heavy elemental enrichment of 100 compared with solar abundances \cite{grevesse1998standard}, which is arbitrary but higher only by a factor of 2 than the carbon enrichment of Uranus and Neptune (\cite{hersant2004enrichment} and references therein). Consequently we obtained a high abundance of \ce{CO2}.

\ce{CO2} has two routes to photodissociate :
\begin{align*}
\ce{CO2} + h\nu \ce{-> CO + O(^3P)} \qquad J4\\
\ce{CO2} + h\nu \ce{-> CO + O(^1D)} \qquad J5
\end{align*}
Depending on the energy of the photons, one route is favored over the other. The quantum yield used for these two photolyses, $q_4(\lambda)$ and $q_5(\lambda)$, are presented in Table~\ref{tab:q} \cite{huebner1992solar}. We assume that they remain the same at high temperature.

\begin{table}[!h]
\begin{center}\begin{tabular}{ll}
\hline
\hline
Quantum yield & Values [wavelength range] \\
\hline
$q_4(\lambda)$ & 1 [167-227]  \\
$q_5(\lambda)$ & variable [50-107] ; 1 [108-166] \\
\hline
\end{tabular}\end{center}
\caption{Quantum yields for the photodissociations of \ce{CO2}}\label{tab:q}
\end{table}

First, we find the steady-state composition of these atmospheres using the absorption cross sections available in the literature, which means at ambient temperature. Then, we replace the "ambient cross section" of \ce{CO2} ($\sigma_{\ce{CO2}}$(300~K)) by the cross section measured at 550~K, between 115 and 200 nm($\sigma_{\ce{CO2}}$(550~K)). For wavelengths between 200 and 230 nm, we use Eq.\ref{eq:formule} to determine $\sigma_{\ce{CO2}}$(550~K).\\

\begin{figure*}[!ht]
\includegraphics[height=0.7\textheight]{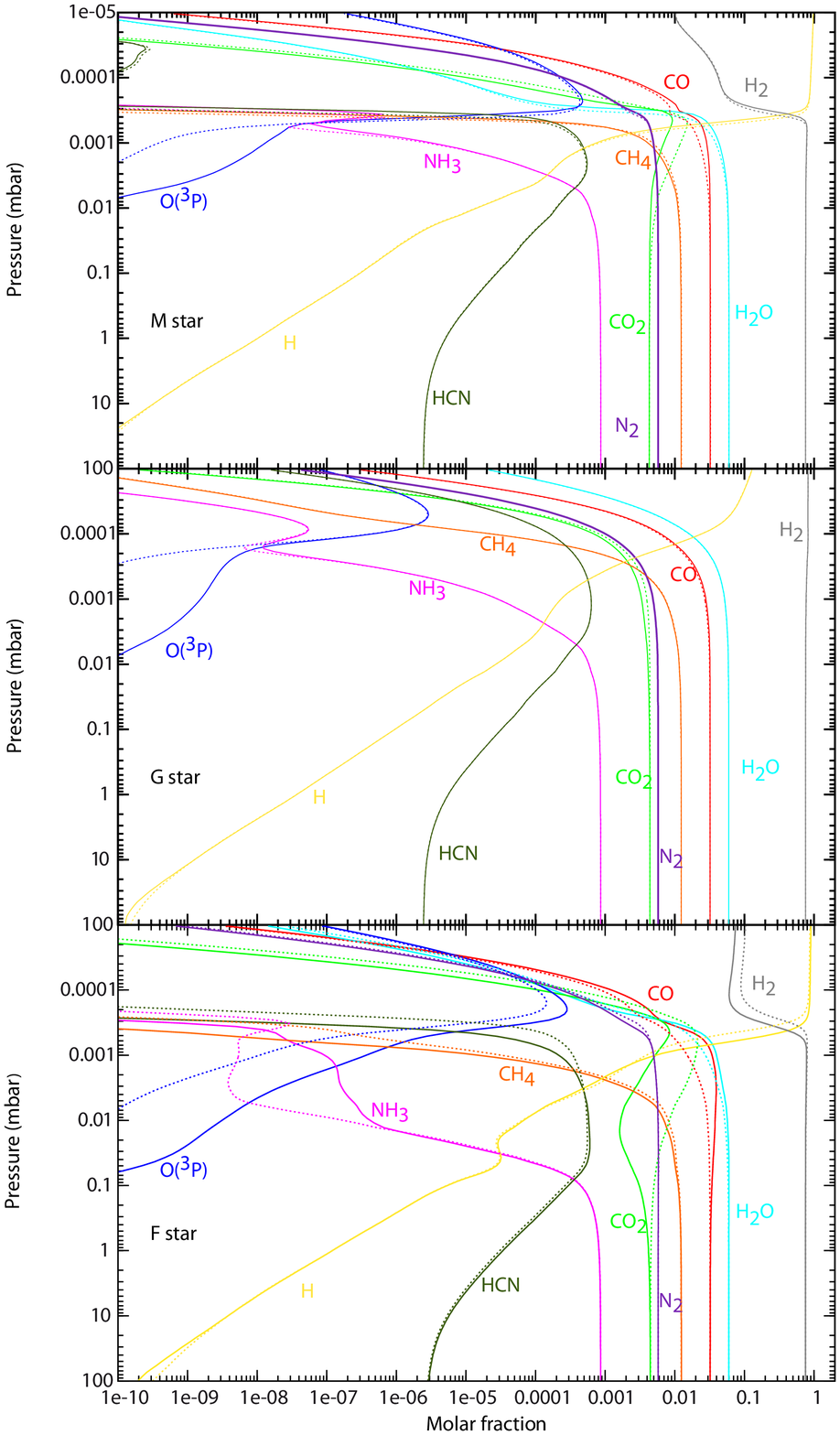}
\includegraphics[height=0.7\textheight]{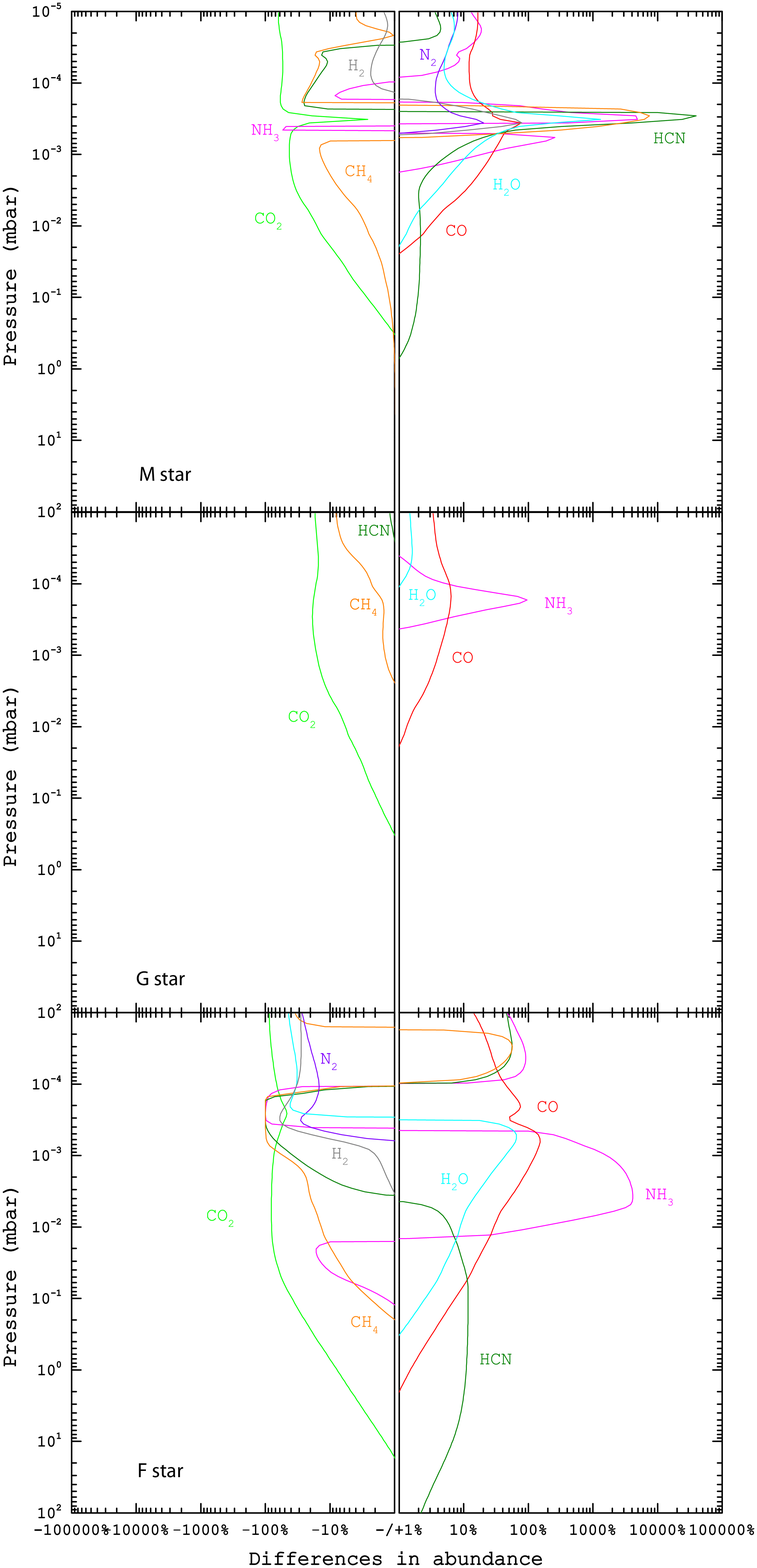}
\caption{\textit{Left:} Comparison of the atmospheric composition of GJ 436b using $\sigma_{\ce{CO2}}$(300~K) (dotted line) and $\sigma_{\ce{CO2}}$(550~K) (full line) when the planet orbits an M star (top), a G star (middle) and an F star (bottom). \textit{Right:} Differences in abundances (in \%) between the results obtained with $\sigma_{\ce{CO2}}$(300~K) and $\sigma_{\ce{CO2}}$(550~K) for species that have an abundance superior to $10^{-10}$, for an M star (top), a G star (middle), and an F star (bottom).}
\label{fig:GJ436b}
\end{figure*}

As we see in Fig.~\ref{fig:GJ436b} (left), changing the absorption cross section of \ce{CO2} has consequences on the abundances of some species. For instance, when considering an M star, the abundance of \ce{CO2} at $5\times10^{-4}$ mbar reduces by 45\% when we use $\sigma_{\ce{CO2}}$(550~K) instead of $\sigma_{\ce{CO2}}$(300~K). The compounds O($^3$P), CO, and \ce{NH3} are also affected by the change of $\sigma_{\ce{CO2}}(T)$. We see that other species are also affected by the change of \ce{CO2} absorption cross section, such as \ce{NH3}, \ce{CH4}, HCN, \ce{H2}, and \ce{N2}. To see these variations, we represented in Fig.~\ref{fig:GJ436b} (right) the differences of abundances of the major species (i.e., for species with an abundance superior to $10^{-10}$) between the two models. We see that the differences in abundances can reach almost $10^{5}$\%, even for species which are not directly linked to $\sigma_{\ce{CO2}}(T)$, such as HCN in the case of the M star.

These differences are easily comprehensible. We chose the case of \ce{NH3} to illustrate it. At high temperature, the absorption cross section of \ce{CO2} is higher around 120~nm and for wavelengths superior to 150~nm, so \ce{CO2} absorbs more UV flux than with the ambient cross section. Consequently, more \ce{CO2} is photolysed. The UV photons that are now absorbed by \ce{CO2} were absorbed by \ce{NH3} when using $\sigma_{\ce{CO2}}$(300~K). Now \ce{NH3} absorbs fewer UV photons, so less is destroyed. This can be generalized to the other species. Indeed, the general behavior of these curves can be explained in terms of the various opacity sources that peak at slightly different altitudes. Because of the competition among the different opacity sources in the atmosphere, all the species absorbing in the same range of wavelength as \ce{CO2} are affected.

It is with the G star that the difference of composition is the least important when we change $\sigma_{\ce{CO2}}$. This is quite normal because the flux received by the planet before 200~nm is lower than with the M star and the F star (see Fig.~\ref{fig:GJ436b}). On the contrary, the difference of composition is the most important in the case F star, because it is the most important flux between 115 and 230 nm.

A discussion about the loss rates of \ce{CO2} in the different cases studied is in \cite{venot2013}.

\section{Conclusion}

We measured absorption cross sections of carbon dioxide at high temperatures for the first time, in the range 115-230~nm.
Between 115 and 200~nm, we measured $\sigma_{\ce{CO2}}(\lambda, T)$ at four temperatures: 300, 410, 480, and 550~K. For longer wavelengths, we made measurements at the following temperatures: 465, 510, 560, 610, 655, 750, and 800~K.  For $\lambda >$ 160~nm, we clearly see that the absorption cross section increases with the temperature. Thanks to the quasi-linear variation of $\ln(\sigma_{\ce{CO2}}(\lambda, T) \times \frac{1}{Q_v(T)})$ after 170~nm, we parametrize the variation of  $\ln(\sigma_{\ce{CO2}}(\lambda, T) \times \frac{1}{Q_v(T)})$ with a linear regression which allows us to calculate $\sigma_{\ce{CO2}}(\lambda, T)$ at any temperature in the range 170-230~nm.
As we show for GJ~436b, these new data have a considerable influence on the loss rate of \ce{CO2} \cite{venot2013}, and on the atmospheric composition of exoplanets that possess high atmospheric temperatures. Placing a hot Neptune around different stars (M, G, and F), we find that  the F star is the star for which the change of absorption cross section has the most influence.\\

To model hot exoplanets, we recommend using cross sections relevant to the atmospheric temperature when available, or at least, as close as possible to the atmospheric temperature. Carbon dioxide is not the only absorbing species of exoplanet atmospheres. 
The influence of the absorption cross section of \ce{CO2} on the atmospheric composition of GJ~436b is only illustrative because the photochemistry results from the fact that species shield each other according to their abundances and their cross sections. We expect that the effect of $\sigma_{\ce{CO2}}(\lambda, T)$ will be more important on other types of atmospheres, in particular \ce{CO2}-rich atmospheres. But the real impact of the temperature dependence of $\sigma_{\ce{CO2}}(\lambda, T)$ can be evaluated only by taking into account the temperature dependence of all the other cross sections. Here, we simply show that it is necessary to establish this dependence for all species that absorb UV radiation. This work on \ce{CO2} is a first step towards this goal. Because \cite{venot2012} show that \ce{NH3} is an important absorber around 200 nm and that it absorbs UV flux very deep in the atmosphere (in pressure regions that can be probed with observations), we plan to measure the absorption cross section of this molecule at temperatures higher than 300~K. Finally, a great deal of work remains to be done in this area which is essential for the photochemical modeling of hot exoplanet atmospheres, whether terrestrial or gaseous. 



\bibliography{bib_article}

\end{document}